\documentclass[showpacs,aps,twocolumn,nofootinbib,preprintnumbers,citeautoscript]{revtex4-1}
\pdfoutput=1
\usepackage{graphicx}
\usepackage{epsfig,amsmath,amsthm}
\usepackage{subfig}
\newcommand{\f}{\begin{equation}}
\newcommand{\ff}{\end{equation}}
\newcommand{\fa}{\begin{eqnarray}}
\newcommand{\ffa}{\end{eqnarray}}

\begin{document}
\title{Metal-insulator Transition by Holographic Charge Density Waves}
\author{Yi Ling $^{1,2,3}$}
\author{Chao Niu $^{1}$}
\author{Jian-Pin Wu $^{4}$}
\author{Zhuo-Yu Xian $^{1}$}
\author{Hongbao Zhang $^{5}$}
\affiliation{$^1$ Institute of High Energy Physics, Chinese
Academy of Sciences, Beijing 100049, China\\
$^2$ Center for Relativistic Astrophysics and High Energy Physics,
Department of Physics, Nanchang University,\\
Nanchang 330031, China\\
$^3$ State Key Laboratory of Theoretical Physics, Institute of
Theoretical Physics, Chinese Academy of Sciences, Beijing 100190,
China\\
$^4$ Department of Physics, School of Mathematics and Physics,
Bohai University, Jinzhou 121013, China\\
$^5$ Theoretische Natuurkunde, Vrije Universiteit Brussel,
and The International Solvay Institutes,\\
Pleinlaan 2, B-1050 Brussels, Belgium
}
\begin{abstract}
We construct a gravity dual for charge density waves (CDW) in
which the translational symmetry along one spatial direction is
spontaneously broken.  Our linear perturbation calculation on the gravity side produces the frequency dependence of
the optical conductivity, which exhibits the two familiar features of CDW, namely the pinned collective mode and gapped single-particle excitation. These two features indicate that our gravity dual also provides a new mechanism to implement the metal to insulator phase transition by CDW, which is further confirmed by the fact that  d.c. conductivity
decreases with the decreased temperature below the critical temperature.

\end{abstract}
\pacs{11.25.Tq, 04.70.Bw}
\maketitle

{\it Introduction.}--In recent years the holographic correspondence between a
gravitational theory and a quantum field theory in condensed
matter physics has been extensively investigated. In particular,
inspired by the seminal work in
\cite{Gubser:2008px,Hartnoll:2008vx}, more and more evidences have
been accumulated in favor of the consensus that many phenomena related to strongly coupled systems may have a
dual description on the gravity side.  In this Letter, we shall
offer a holographic mechanism to implement the metal-insulator phase transition by CDW.

CDW is a novel ground state of the coupled
electron-phonon system, which is
characterized by a collective mode formed by electron-hole pairs
with a wave vector $q=2k_F$ and a gap in the single-particle
excitation spectrum\cite{Gruner:1988zz}. Ideally, this collective mode would sit at zero frequency, leading to a supercurrent. But due to the inevitable interaction between CDW and the underlying background, this collective mode is generically shifted to a finite resonance frequency. This pinning effect together with the gapped single-particle excitation suggests a mechanism to induce metal-insulator phase transition by CDW. With this in mind, it is highly possible to provide a holographic realization of metal-insulator transition once a holographic CDW is implemented.

It should be stressed that the generation of CDW in condensed
matter corresponds to the {\it spontaneous} breaking of
translational symmetry. Therefore, to implement a holographic
description of CDW, it is essential to introduce some mechanism of
the modulated instability of the bulk geometry which is usually of
spatial homogeneity. This issue has been addressed recently and
the examples of spatially modulated unstable modes have been
presented\cite{Ooguri:2010xs,Nele,Donos:2011bh,Lippert1,Lippert2,Erdmenger,Lippert3,Rozali:2012es,Donos:2013wia,Withers:2013loa,Withers:2013kva,Rozali:2013ama,Donos:2013gda}.
However, until now the study on the dynamics of CDW by holography is still absent. Thus it is unclear not only whether
such holographic CDW reproduce
the observed fundamental features of ordinary CDW
in experiments but also whether the corresponding CDW phase transition
is accompanied by the metal-insulator transition. We shall make a first attempt to address these issues and provide
an affirmative answer to both of these questions by investigating the optical
conductivity of holographic CDW in a striped black hole
background.

{\it Holographic setup and background solutions.}--Our model consists of
gravity coupled with two gauge fields plus a dilaton field in four dimensions
\begin{eqnarray}
S&=&\frac{1}{2\kappa^2}\int
d^4x\sqrt{-g}[R+\frac{1}{L^2}-\frac{1}{4}t(\Phi)F^{\mu\nu}F_{\mu\nu}
-\frac{1}{4}G^{\mu\nu}G_{\mu\nu}\nonumber\\
&&-\frac{1}{2}(\partial_\mu\Phi\partial^\mu\Phi+m^2\Phi^2)
-\frac{1}{2}u(\Phi)F^{\mu\nu}G_{\mu\nu}], \label{eq:action}
\end{eqnarray}
where $F=\mathrm{d}A$, $G=\mathrm{d}B$, $t(\Phi)=1-\frac{\beta}{2}L^2\Phi^2$,
and $u(\Phi)=\frac{\gamma}{\sqrt{2}}L\Phi$. The first gauge field $A$
is introduced to form an AdS-RN black hole background with finite
temperature and non-vanishing chemical potential, while the second
gauge field $B$ as well as the dilaton field will be responsible for the instability of the
background and the CDW phase will be associated with this second
$U(1)$ symmetry\cite{Donos:2013gda}.
Below we shall set the AdS radius
$l^2=6L^2=\frac{1}{4}$ and $m^2=-\frac{2}{l^2}=-8$. In addition,  $\frac{l^2}{2\kappa^2}\gg1$ is required such that classical gravity is reliable, which corresponds to the large $N$ limit of the dual field theory. Thus in our setup the remaining adjustable parameters
are $\beta$ and $\gamma$.

Obviously, in the case of $\Phi=0$ and $B=0$, the equations of
motion always allow the electric AdS-RN black hole solution
\fa ds^2={1\over
z^2}\left(-(1-z)f(z)dt^2+\frac{dz^2}{(1-z)f(z)}+dx^2+dy^2\right)\nonumber\\
\label{metric}\ffa
with
\fa f(z)=4(1+z+z^2-\frac{z^3\mu^2}{16}), \ \ \ \ A_t=\mu(1-z).
\ffa
We will consider the dual field theory in a grand
canonical system, hence we will use the chemical potential $\mu$ as the
unit for the system. In this coordinate system, the black hole
horizon is located at $z=1$ and the $AdS_4$ boundary is at $z=0$.
The Hawking temperature of the black hole is $
T/\mu=(48-\mu^2)/(16\pi\mu)$. The zero temperature limit is
reached when $\mu=4\sqrt{3}$. However,  the linear perturbation analysis shows that at low temperature such a black hole will be unstable against the striped phase\cite{Donos:2013gda}. To obtain such a resultant striped solution by solving the fully non-linear bulk dynamics numerically,
we assume the following ansatz for the background fields
\begin{eqnarray}
ds^2&=&{1\over
z^2}[-(1-z)f(z)Qdt^2+\frac{Sdz^2}{(1-z)f(z)}+Vdy^2+\nonumber\\
&&T(dx+z^2Udz)^2],\nonumber\\
A&=&\mu(1-z)\psi dt,\nonumber\\
B&=&(1-z)\chi dt,\nonumber\\
\Phi &=& z \phi,
\end{eqnarray}
where the eight variables involved in the ansatz are functions of $x$ and $z$. In order to have a
spontaneous breaking of the translational symmetry in the dual field
theory,  the
following Dirichlet boundary conditions are imposed \fa
&Q[x,0]=S[x,0]=T[x,0]=V[x,0]=\psi[x,0]=1,\nonumber\\
&U[x,0]=\chi[x,0]=\phi[x,0]=0. \ffa Furthermore, we impose the
regularity conditions at the horizon such that all the functions
have a Taylor expansion in powers of $(1-z)$. Now the equations of motion reduce to eight partial differential
equations with respect to $x$ and $z$. We solve them numerically
with the Einstein-DeTurck method, which has been employed to look
for static solutions to Einstein equations\cite{Headrick:2009pv,Horowitz:2012ky,Horowitz:2012gs,Horowitz:2013jaa,Ling:2013aya,Ling:2013nxa}.
We demonstrate the
relevant result below, where we focus solely on the case of $\beta=-138$ and $\gamma=17.1$.

\begin{figure}
\center{
\includegraphics[scale=0.2]{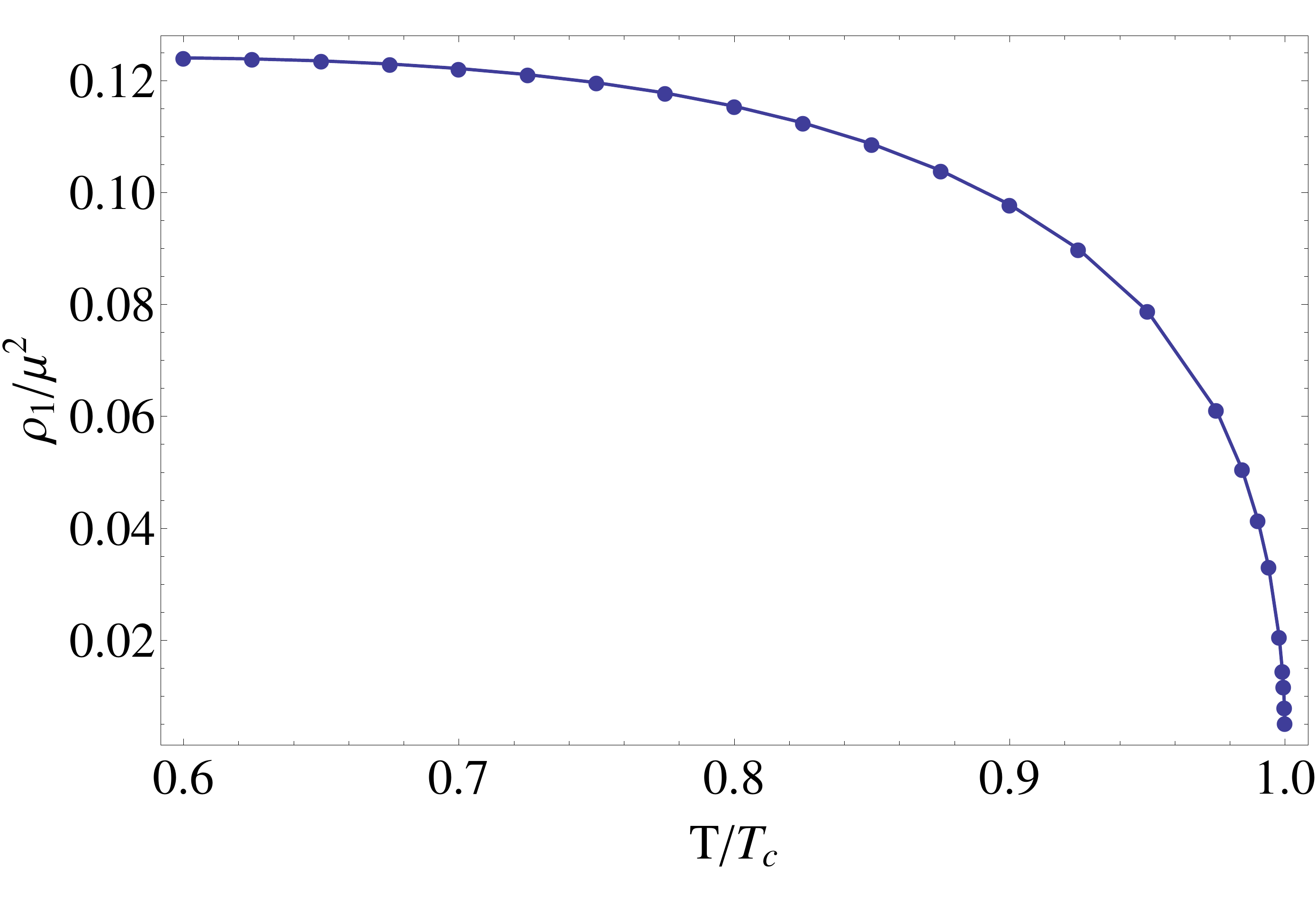}\hspace{0.1cm}
\caption{\label{mode}The first mode of CDW as a function of temperature.} }
\end{figure}

\begin{figure}
\center{
\includegraphics[scale=0.14]{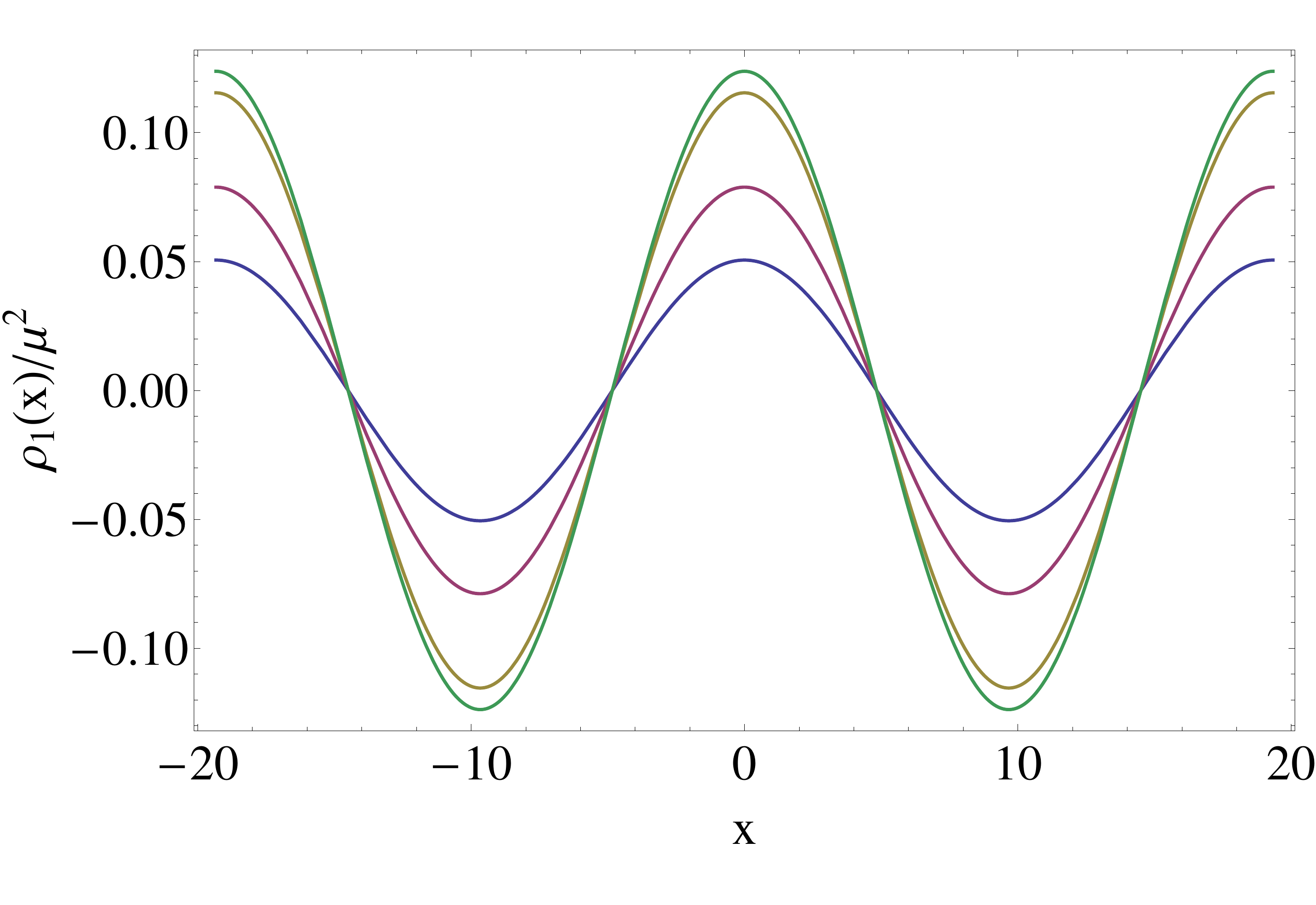}\hspace{0.1cm}
\includegraphics[scale=0.14]{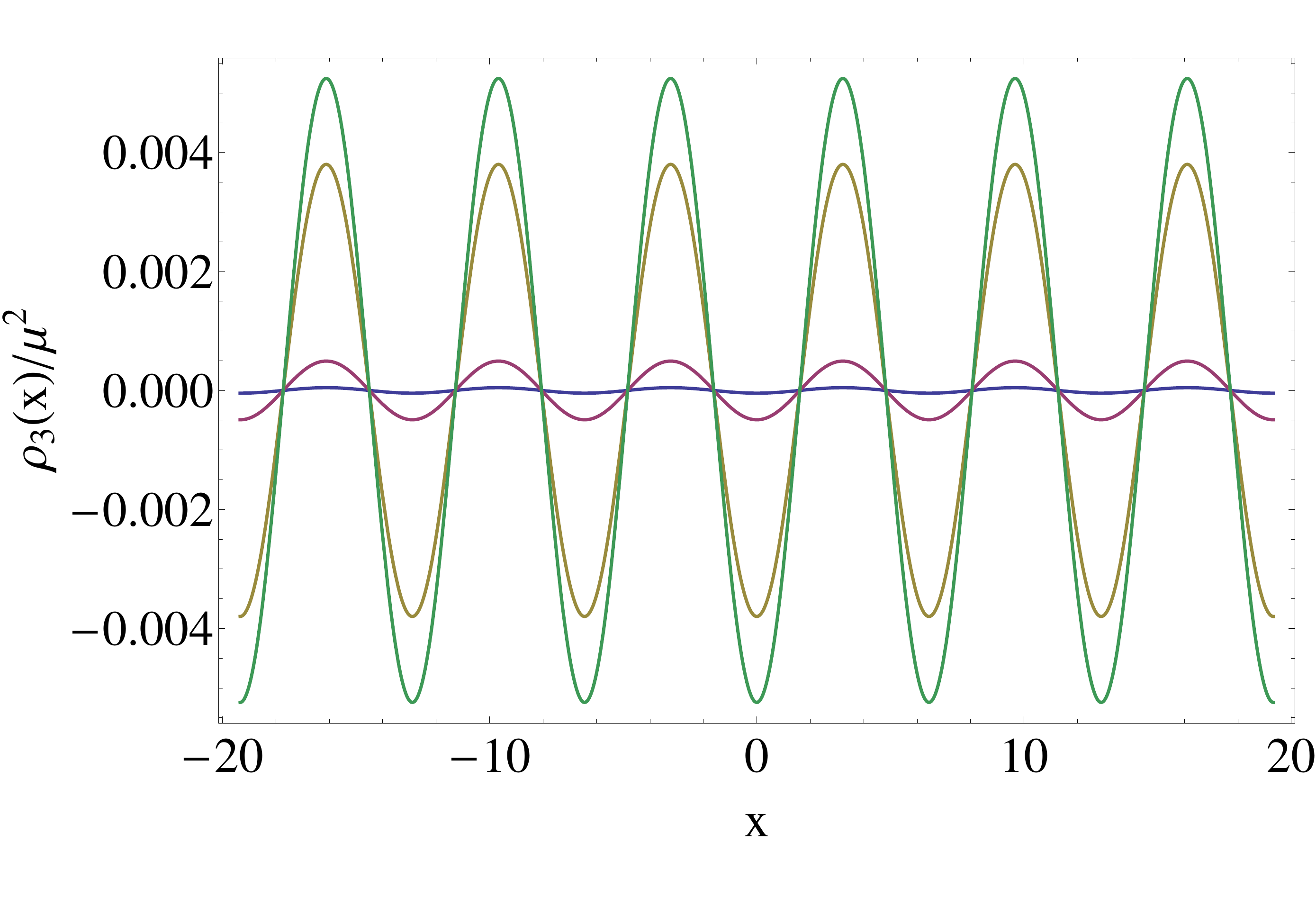}\hspace{0.1cm}
\caption{\label{mode2}The first and third modes of CDW for $T/T_c=0.6, 0.8, 0.95,0.98$ from top to down.} }
\end{figure}

The corresponding critical temperature for the phase transition to the striped phase is
about $T_c=0.078\mu$ and the critical momentum mode in the $x$ direction is  given by $k_c=0.325\mu$.
The onset of CDW can be read
off explicitly from the component of the gauge field $B_t$\cite{notation}
 \fa
&B_t=-\rho(x)z+O(z^2),\nonumber\\
&\rho(x)=\rho_0+\rho_1\cos[k_cx]+...+\rho_n\cos[nk_cx]+... .\ffa
We find that the coefficients of even orders in numerical
solutions vanishes, such that the charge density for CDW
has the form $\rho(x)=\rho_1\cos[k_cx]+\rho_3\cos[3k_cx]+...$. As shown in Fig.\ref{mode}, $\rho_1$ can serve as the order parameter of our system to characterize the phase transition to CDW as it should be the case. Its condensation behavior near the critical
temperature indicates that the system undergoes a second order
phase transition to CDW. In
Fig.\ref{mode2} we have plotted the charge density associated with
$\rho_1$ and $\rho_3$ at various temperatures, respectively. From this figure we
notice that near the critical temperature the sub-leading term
$\rho_3$ is tiny comparing with the leading term $\rho_1$ and can
be neglected, while as the temperature goes down, its contribution
becomes important.

\begin{figure}
\center{
\includegraphics[scale=0.14]{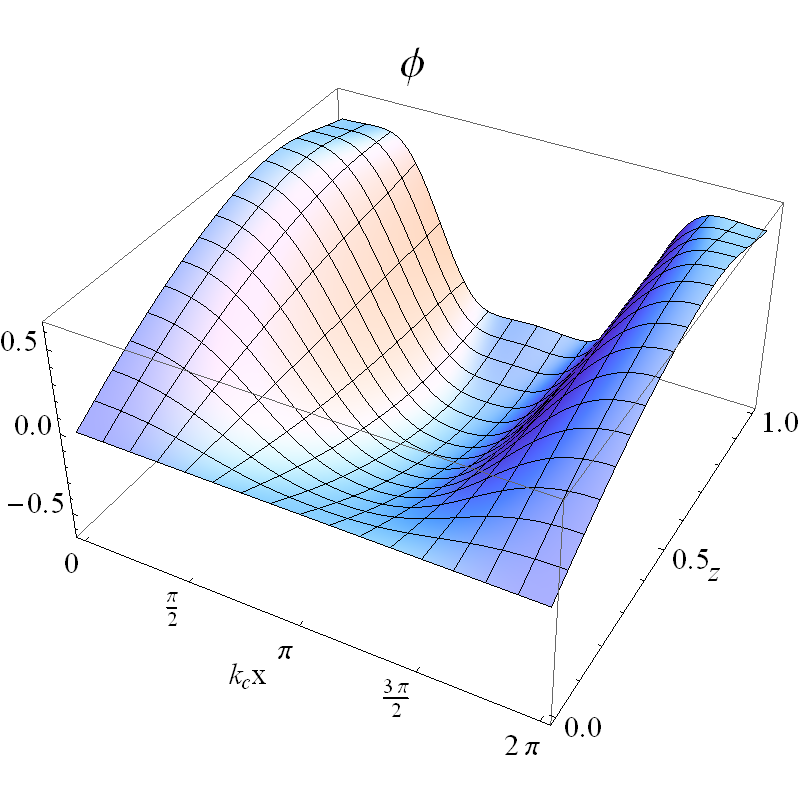}\hspace{0.1cm}
\includegraphics[scale=0.14]{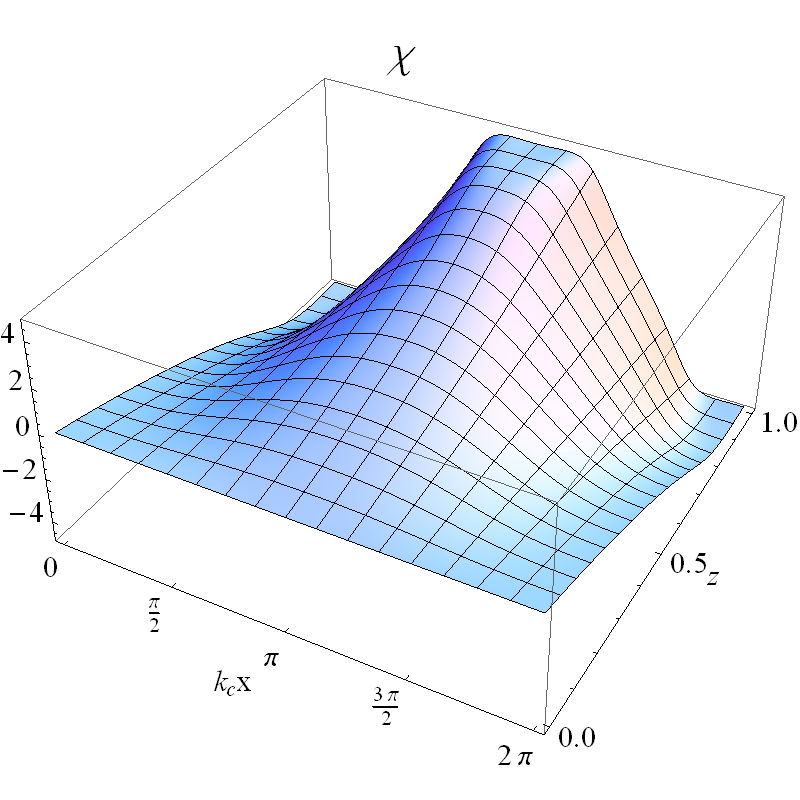}\hspace{0.1cm}
\caption{\label{solu}Solutions of the scalar field and the time
component of the gauge field $\chi$ for $T=0.8T_c$.} }
\end{figure}

In
Fig.\ref{solu} we plot the solutions of the scalar $\phi$ and the
time component of the gauge field $\chi$ at the temperature
$T=0.8T_c$. Note that the striped profile increases when one goes deeper into the horizon, which is consistent with the linear perturbation analysis that such a striped phase is triggered by the instability of near horizon geometry $AdS_2\times R^2$ of extremal AdS-RN black hole\cite{Donos:2013gda}. With this relevant striped deformation, the IR metallic fixed point characterized by $AdS_2\times R^2$ is  driven to another fixed point. As we shall show in the next section, this resultant fixed point corresponds to an insulating phase.


{\it Optical conductivity of holographic CDW and metal-insulator transition.}--Now we turn to study the dynamics of holographic CDW by computing
the optical conductivity as a function of
frequency. To this end, we separate the variables into the background part and
fluctuation part as
\begin{equation}
g_{\mu\nu}=\bar{g}_{\mu\nu}+h_{\mu\nu}, A_{\mu}=
\bar{A}_{\mu}+a_{\mu}, B_{\mu}= \bar{B}_{\mu}+b_{\mu},
\Phi=\bar{\Phi}+\varphi.
\end{equation}
 We assume that the
fluctuations of all the fields have a time dependent form as
$e^{-i\omega t}$ but
independent of the coordinate $y$.
To solve the fluctuation
equations, gauge conditions must be imposed for gravity and
two gauge fields. Here we choose the de Donder gauge and Lorentz
gauge condition for them, respectively \fa \bar{\nabla}^\mu
\hat{h}_{\mu\nu} =0,\qquad \bar{\nabla}^\mu a_\mu=0, \qquad
\bar{\nabla}^\mu b_\mu=0\ffa where
$\hat{h}_{\mu\nu}=h_{\mu\nu}-h\bar{g}_{\mu\nu}/2$ is the
trace-reversed metric perturbation.

As usual, we adopt ingoing wave boundary conditions at the
horizon. While at our AdS boundary $z=0$, we consider the following consistent
boundary condition with
\fa
&b_x(x,0)=1,\ \ a_x(x,0)=\frac{\partial_z\chi(x,0)}{\mu(1-\partial_z\psi(x,0))} \nonumber\\
&others(x,0)=0.
\ffa
Then by holography, we can extract the homogeneous part of optical conductivity, the quantity we are interested in. Namely, given that $b_x=(1+j_x(x)z+...)e^{-i\omega t}$ by solving the fluctuation equations, the
conductivity associated with the second gauge field reads
$\sigma(\omega/\mu)=\frac{4j_x^{(0)}}{i\omega}$, in which a factor four comes
from the unusual asymptotic form of the metric in
Eq.(\ref{metric}).

\begin{figure}
\center{
\includegraphics[scale=0.28]{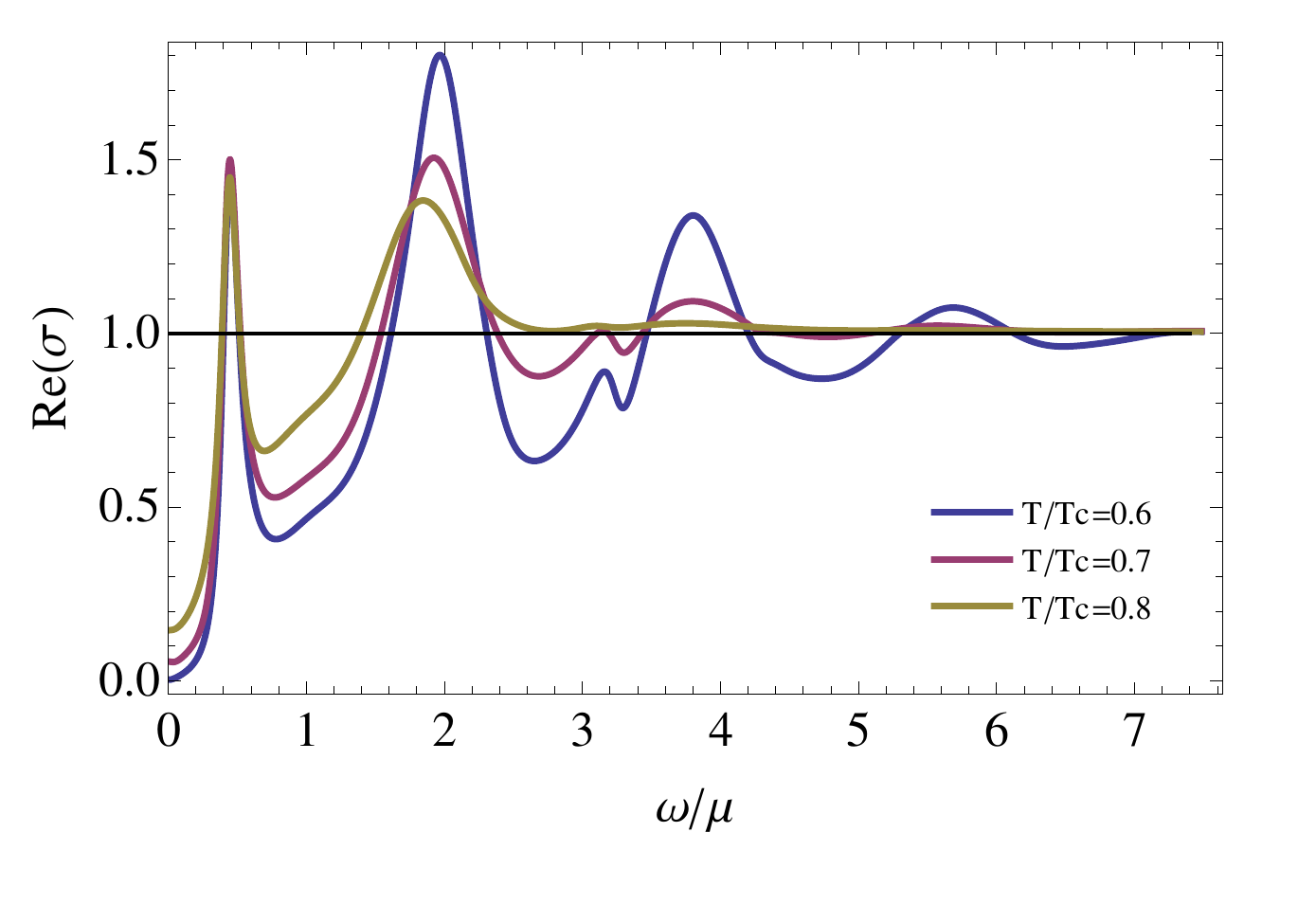}\hspace{0.1cm}
\includegraphics[scale=0.28]{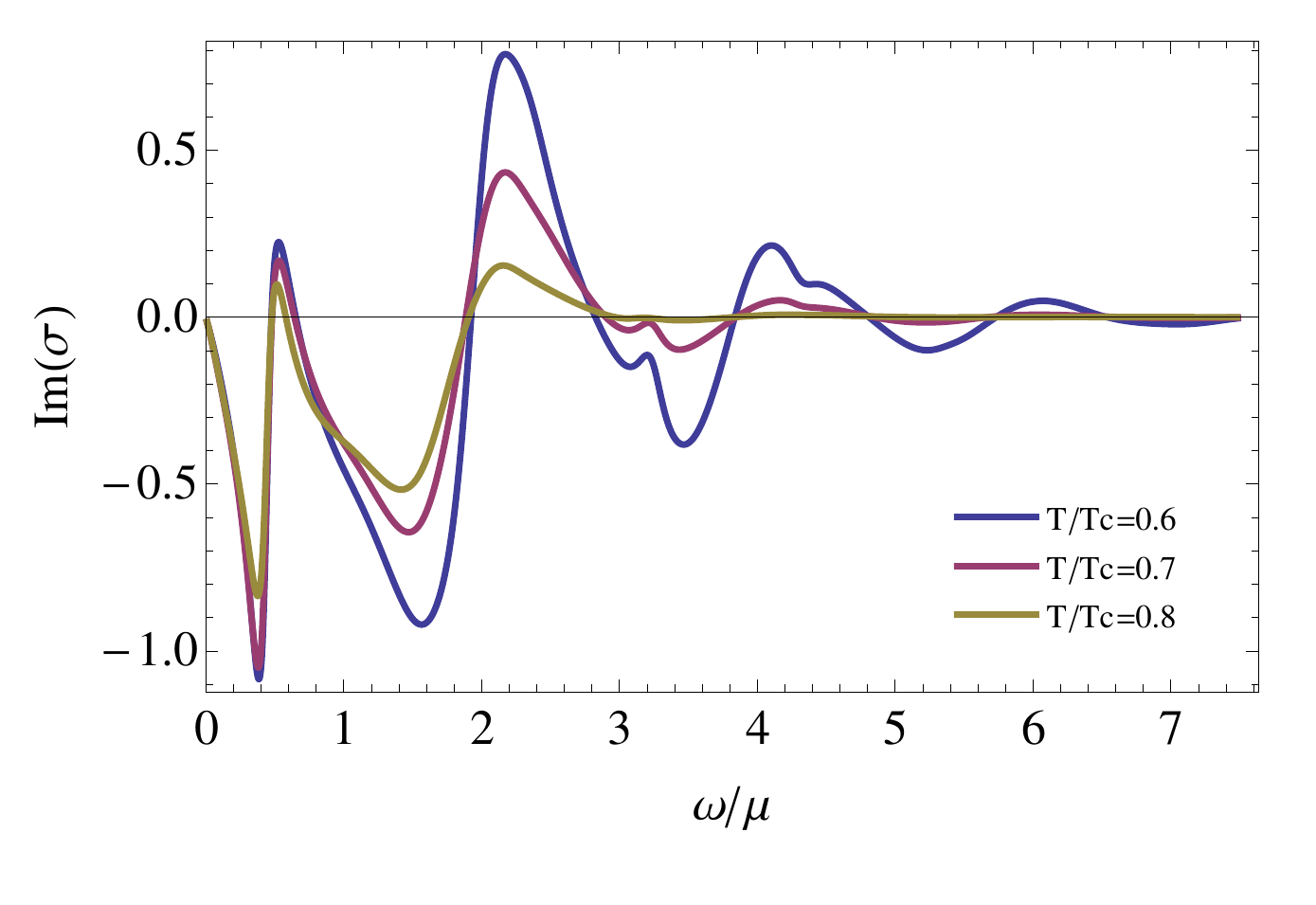}\hspace{0.1cm}
\caption{\label{cond}The optical conductivity for CDW, where the black horizontal line denotes the corresponding optical conductivity for AdS-RN black hole associated with the second gauge field.} }
\end{figure}

One typical plotting for the real and
imaginary parts of the optical conductivity at various
temperatures is shown in Fig.\ref{cond}. Two fundamental features of CDW
are observed. One is the pinned collective mode, which is reflected as the
first peak appearing in the real part of the conductivity. The
second is the gapped single-particle excitation, which corresponds
to the occurrence of the second peak in the real part of the
conductivity.

\begin{figure}
\center{
\includegraphics[scale=0.14]{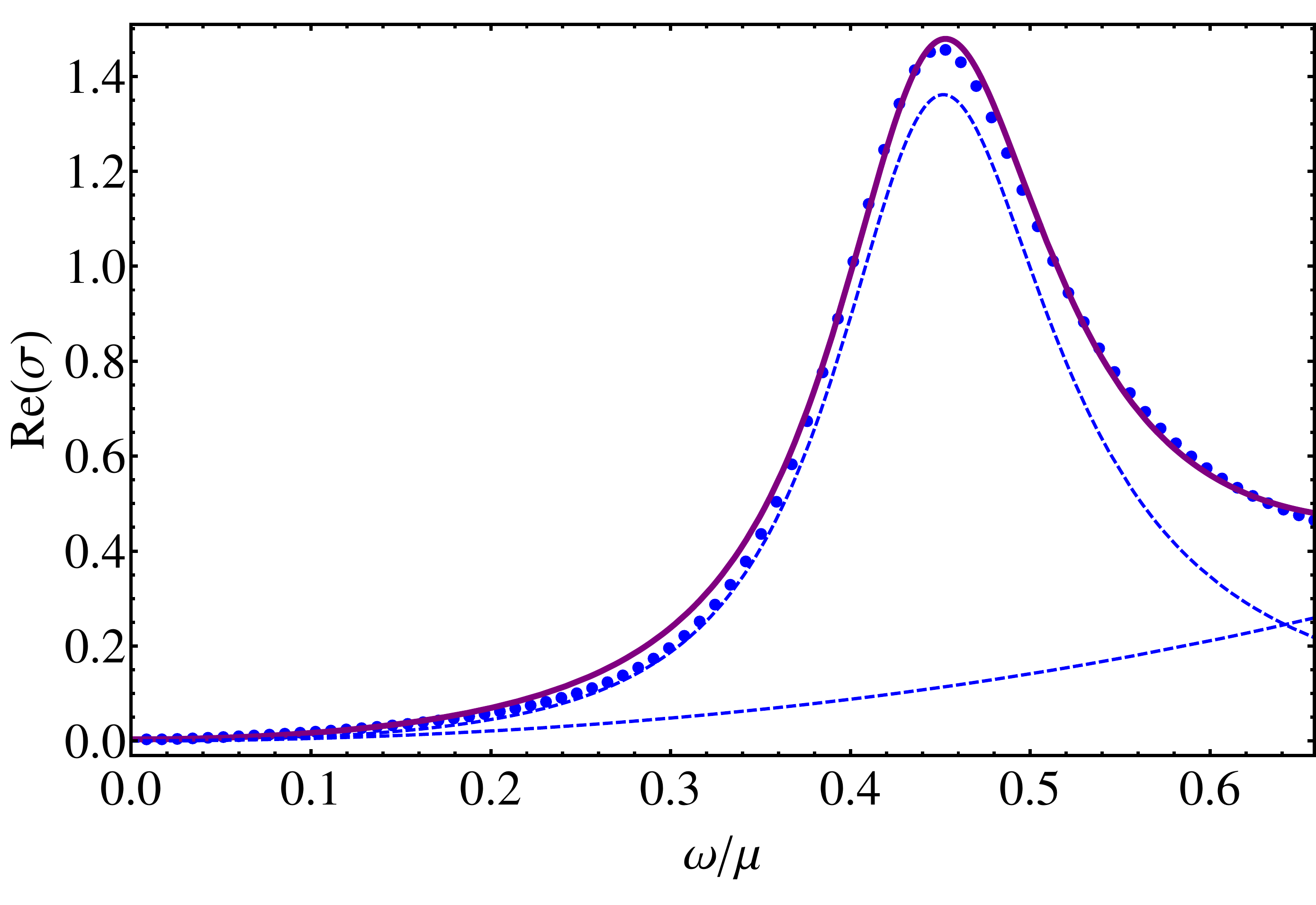}\hspace{0.1cm}
\includegraphics[scale=0.14]{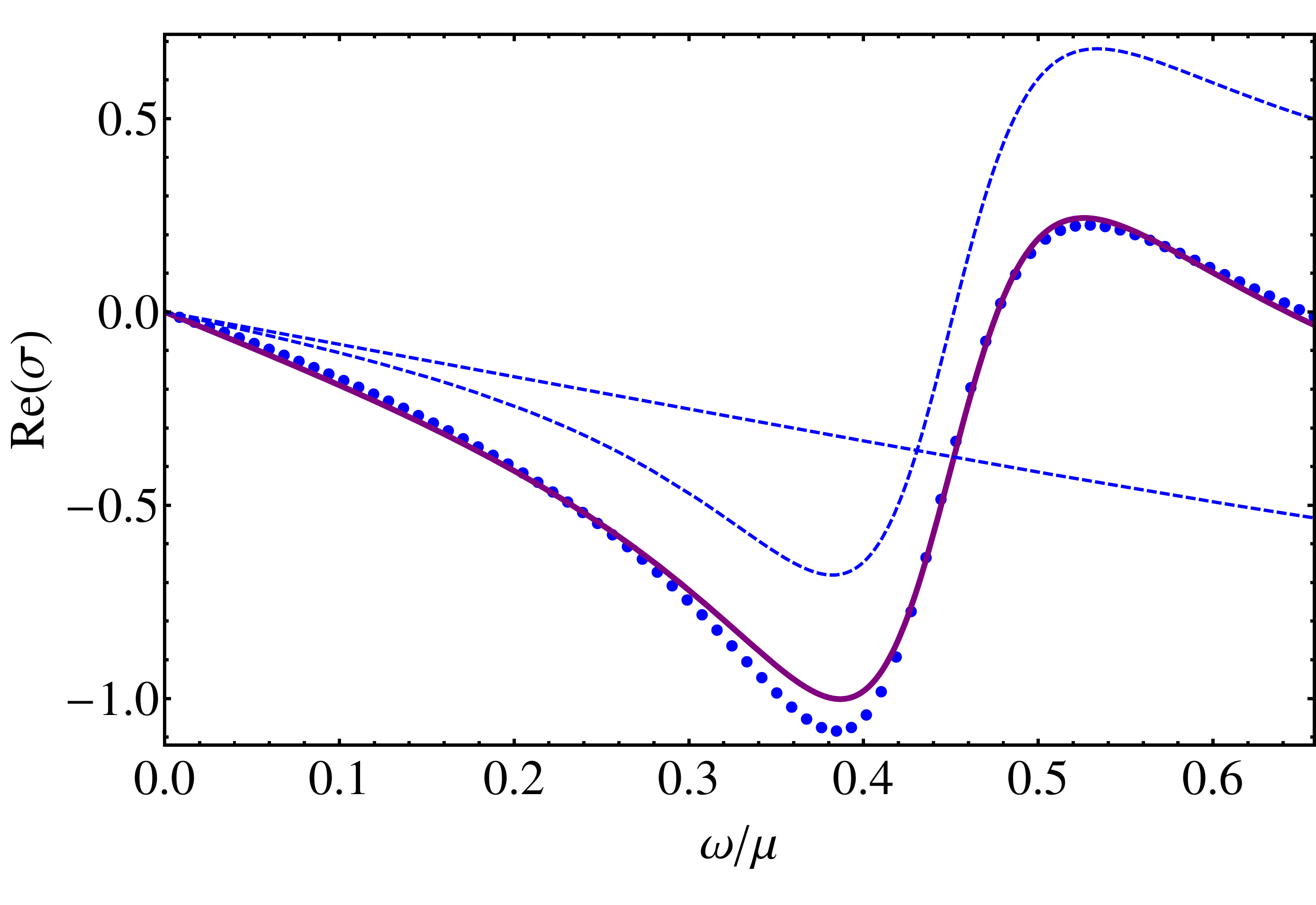}\hspace{0.1cm}
\caption{\label{fit} The fit of
optical conductivity with two Lorentz oscillators in the low frequency regime for $T=0.6T_c$.  The contributions from the individual
oscillator are also plotted with dashed lines.}}
\end{figure}

The pinning is a common phenomenon for CDW on
account of the various interactions with the other components of the system, which can be described by a damped harmonic
oscillator with Lorentz resonance \fa \sigma_{CDW}(\omega)=\frac{K
\tau}{1-i\omega\tau(1-\omega_0^2/ \omega^2)},\label{ccdw} \ffa
where $\tau$ is the relaxation constant, $K$ is proportional to
the number density of CDW, and $\omega_0$ is the average pinning resonance frequency\cite{Gruner:1988zz}.
This formula has been widely
employed in the analysis of CDW optical response experiments. Here we also use it to fit our data.
In consistent with the fact that our
holographic CDW is always generated with multiple wave vectors,
we find in general our data in the low frequency region of the
conductivity can be well fit with multiple Lorentz oscillators. In
particular, as the temperature is not quite low, for instance
$T\geq 0.6T_c$, it can be fit with only two oscillators,  namely $
\sigma_{tot}(\omega)=\sigma_{CDW1}(\omega)+\sigma_{CDW2}(\omega)\label{ccdw}
$, because in this case the contribution from those CDW with higher wave vectors is negligible. Fig.\ref{fit} is such a fit to this formula for  $T/T_c=0.6$. The
parameters in the Lorentz formula for various temperatures are listed
in Table \ref{parameters}.

\begin{table}[ht]
\begin{center}

\begin{tabular}{|c|c|c|c|c|c|c|}
\hline
$T/T_c$&$K_1/\mu$&$\tau_1\mu$&$\omega_{01}/\mu$&$K_2/\mu$&$\tau_2\mu$&$\omega_{02}/\mu$\\
\hline
$0.6$&$0.207$&$6.593$&$0.452$&$2.225$&$0.609$&$1.629$\\
\hline
$0.7$&$0.207$&$6.327$&$0.449$&$2.512$&$0.498$&$1.653$\\
\hline
$0.8$&$0.188$&$5.870$&$0.442$&$1.587$&$0.802$&$1.354$\\
\hline
\end{tabular}
\caption{\label{parameters} The fit parameters in
Lorentz formula at various temperatures.}

\end{center}
\end{table}

 Although this pinned collective mode is gapless, our
single-particle excitation is gapped, as clearly evident from Fig.\ref{cond}. In particular,
the magnitude of gap is estimated as $2\Delta/T_c\approx 20.51$ by locating the
position of the second minimum in the imaginary part of the
conductivity, which is obviously much larger than the mean-field BCS value $2\Delta/T_c\approx 3.52$. This large gap ratio associated with this gap is indicative of a strongly coupled CDW phase transition in our system as it should be the case by holography. On the other hand, remarkably, this large gap ratio turns out to be comparable to some CDW
materials. For example, it follows from the experimental data on the optical conductivity that the gap ratio is given by $2\Delta/T_c\approx15.80$ for the single crystalline TbTe3 compound\cite{Chen}.

Now with the spectral weight transferred to our pinned collective mode and gapped single-particle excitation, the resultant CDW can be identified as an insulator, as further evidenced by the decreasing behavior of conductivity at zero frequency with the decreased temperature. Thus our holographic CDW provides an alternative implementation of metal-insulator transition\cite{DonosHartnoll}.

We conclude this section by a remark. In the large frequency regime, as the
temperature goes down, the contribution from higher order CDW
will become relevant such that more peaks and gaps emerge in
this regime, which have been observed in
Fig.\ref{cond} when $T\leq 0.7T_c$. Describing these new
resonances quantitatively require one to go to much lower temperature, which involves heavier numerical computation
and is beyond the scope of our Letter.

{\it Discussion.}--We have constructed a new type of striped black
hole solutions which is characterized by the condensation of
CDW. Two fundamental features of CDW have been precisely reproduced by investigating the optical
conductivity of our holographic CDW. Together with the behavior of d.c. conductivity, we are successfully  led to a new mechanism of the metal-insulator transition by holographic CDW.

In addition, taking into account that the significantly large gap ratio of our holographic CDW is comparable to the experimental data on some CDW materials, our work opens a promising window
for understanding the related phenomena of CDW in condensed matter
physics by holography.

We would like to end this Letter with one important reminder. As mentioned before, we have worked in the large $N$ limit, therefore the dangerous infrared thermal or quantum fluctuations are parametrically suppressed as $1/N$ corrections\cite{Witten,AHI}, which explains how we can have the spontaneous breaking of translational symmetry along one dimension in a 2+1 dimensional system, in apparent contradiction to the Landau-Peierls theorem\cite{Peierls,Landau,BFG}. Note that the two or three dimensional CDW phase is generically robust against fluctuations, thus it is significant to explicitly check whether our main results obtained here are carried over onto higher dimensional holographic CDW, although it should be the case. But the involved numerical computation is extremally non-trivial and expected to be reported in the future.

{\it Acknowledgements.}--We are grateful to Shu Chen, Aristomenis Donos, Xianhui Ge,
Jianlin Luo, Ioannis Papadimitriou, Jorge Santos and Nanlin Wang
for helpful discussions. This work is supported by the Natural
Science Foundation of China under Grant Nos.11275208, 11305018 and
11178002. Y.L. also acknowledges the support from Jiangxi young
scientists (JingGang Star) program and 555 talent project of
Jiangxi Province. H.Z. is supported in part by the Belgian Federal
Science Policy Office through the Interuniversity Attraction Pole
P7/37, by FWO-Vlaanderen through the project G.0114.10N as well as
G020714N, and by the Vrije Universiteit Brussel through the
Strategic Research Program ``High-Energy Physics". He is also grateful to the organizers of the holographic inhomogeneities workshop in Amsterdam for their wonderful hospitality and the participants for the helpful discussions related to this work.


\begin{thebibliography}{99}
\bibitem{notation}
In what follows, we shall scale out the prefactor $\frac{1}{2\kappa^2}$ of our charge density and conductivity for the notational convenience. Readers can restore it by multiplying our data with it if necessary.
\end{thebibliography}
\end{document}